\documentclass[conference]{IEEEtran}

\usepackage{lipsum}  
\usepackage[T1]{fontenc}
\usepackage{blindtext}
\usepackage{makecell}
\usepackage{graphicx}
\usepackage{algorithmic}
\usepackage[utf8]{inputenc}
\usepackage{xcolor,colortbl}
\usepackage{enumitem}
\usepackage{booktabs}

\definecolor{mygray}{gray}{0.6}

\usepackage{scalerel}
\usepackage{tikz}
\usetikzlibrary{svg.path}

\definecolor{orcidlogocol}{HTML}{A6CE39}
\tikzset{
  orcidlogo/.pic={
    \fill[orcidlogocol] svg{M256,128c0,70.7-57.3,128-128,128C57.3,256,0,198.7,0,128C0,57.3,57.3,0,128,0C198.7,0,256,57.3,256,128z};
    \fill[white] svg{M86.3,186.2H70.9V79.1h15.4v48.4V186.2z}
                 svg{M108.9,79.1h41.6c39.6,0,57,28.3,57,53.6c0,27.5-21.5,53.6-56.8,53.6h-41.8V79.1z M124.3,172.4h24.5c34.9,0,42.9-26.5,42.9-39.7c0-21.5-13.7-39.7-43.7-39.7h-23.7V172.4z}
                 svg{M88.7,56.8c0,5.5-4.5,10.1-10.1,10.1c-5.6,0-10.1-4.6-10.1-10.1c0-5.6,4.5-10.1,10.1-10.1C84.2,46.7,88.7,51.3,88.7,56.8z};
  }
}

\newcommand\orcidicon[1]{\href{https://orcid.org/#1}{\mbox{\scalerel*{
\begin{tikzpicture}[yscale=-1,transform shape]
\pic{orcidlogo};
\end{tikzpicture}
}{|}}}}

\usepackage{hyperref} 

\begin{document}

\title{Assessing Smart Contracts Security Technical Debts}




\author{\IEEEauthorblockN{1\textsuperscript{st} Sabreen Ahmadjee  \orcidicon{0000-0003-4553-4770}}
\IEEEauthorblockA{\textit{School of Computer Science} \\
\textit{University of Birmingham}\\
Birmingham, UK \\
\textit{College of Computer and Information Systems} \\
\textit{Umm Al-Qura University} \\
Makkah, Saudi Arabia \\
smahmadjee@uqu.edu.sa}
\and
\IEEEauthorblockN{2\textsuperscript{nd} Carlos Mera-G\'omez \orcidicon{0000-0002-7014-1138}} 
\IEEEauthorblockA{\textit{ESPOL Polythecnic University} \\
\textit{Escuela Superior Politécnica del}\\
Litoral, ESPOL, Facultad de Ingeniería\\
en Electricidad y Computación \\
Campus Gustavo Galindo Km 30.5 Vía\\
Perimetral, P.O. Box 09-01-5863\\
Guayaquil, Ecuador\\
cjmera@espol.edu.ec}
\and
\IEEEauthorblockN{3\textsuperscript{rd} Rami Bahsoon}
\IEEEauthorblockA{\textit{School of Computer Science} \\
\textit{University of Birmingham}\\
Birmingham, UK \\
r.bahsoon@cs.bham.ac.uk}
}

\maketitle


\begin{abstract}
Smart contracts are self-enforcing agreements that are employed to exchange assets without the approval of trusted third parties. This feature has encouraged various sectors to make use of smart contracts when transacting. Experience shows that many deployed contracts are vulnerable to exploitation due to their poor design, which allows attackers to steal valuable assets from the involved parties. Therefore, an assessment approach that allows developers to recognise the consequences of deploying vulnerable contracts is needed. In this paper, we propose a debt-aware approach for assessing security design vulnerabilities in smart contracts. Our assessment approach involves two main steps: (i) identification of design vulnerabilities using security analysis techniques and (ii) an estimation of the ramifications of the identified vulnerabilities leveraging the technical debt metaphor, its principal and interest. We use examples of vulnerable contracts to demonstrate the applicability of our approach. The results show that our assessment approach increases the visibility of security design issues. It also allows developers to concentrate on resolving smart contract vulnerabilities through technical debt impact analysis and prioritisation. Developers can use our approach to inform the design of more secure contracts and for reducing unintentional debts caused by a lack of awareness of security issues.

\end{abstract}






\section{Introduction}


A smart contract is a decentralised code agreement designed to impose an automatic negotiation of a series of instructions without requiring approval by a central authority \cite{wang2019blockchain}. Despite the infancy of smart contracts, their emergence has disrupted several sectors, including cryptocurrencies, financial services, insurance, healthcare, and decentralised management, among others \cite{gatteschi2018blockchain}.
The widespread usage of this emerging technology has incentivised attackers to exploit its existing security and privacy challenges. 
Various malicious attacks have been accomplished due to deploying poorly designed or vulnerable smart contracts 






Uploading smart contract to the public blockchain is not free as a specific amount of money is required to be paid for each deployment process \cite{yu2020smart}. Moreover, different from traditional distributed software that can be patched when vulnerabilities are discovered, smart contracts are irreversible and immutable \cite{luu2016making}. These properties mean that once a smart contract is deployed in the chain, it cannot be subsequently modified. Therefore, a timely identification of vulnerability is crucial to save business value \cite{izurieta2018position, izurieta2019leveraging}. In practice, design vulnerabilities do not only originate from 50\% of security issues \cite{santos2017catalog} but they are also the most harmful \cite{sousa2018identifying} and difficult to identify \cite{nord2016can, yavo2016}. Securing smart contracts calls for approaches that can accelerate the identification of root causes of vulnerabilities in the design of such contracts. 


The novel contribution of this paper is a debt-aware approach for assessing security design vulnerabilities in smart contracts. The technical debt metaphor has proven to be effective to measure the impact of security weaknesses exploitation in terms of damage to business value \cite{izurieta2019leveraging}, the ramifications of negative decisions over time, and locate the design root of security vulnerabilities \cite{nord2016can}. Then, we leverage on the metaphor to estimate the monetary cost of redeploying the patched version of the vulnerable contract and the evolution of the debt interest linked to a design vulnerability. Our approach is based on automated analysis tools to discover potential security vulnerabilities caused by design decisions. The evolution of the negative consequences of these security issues in the smart contracts are estimated as an analogy with the concepts of debt principal and interest growth rate.  

This research intends to answer the following research questions (RQ): 
 \textbf{RQ1:} How to identify design vulnerabilities in smart contracts? What are the specific analysis techniques and tools? 
 \textbf{RQ2:} How to quantify the impact of technical debts related to design vulnerabilities in smart contracts?


Although prior works have introduced the idea of technical debt in the context of software security \cite{izurieta2018position, izurieta2019leveraging, nord2016can, rindell2019managing}, to the best of our knowledge, our work is the first to introduce the metaphor to discuss design vulnerabilities that lead to security issues in smart contracts. Moreover, although previous works have surveyed smart contract vulnerabilities \cite{li2020survey, hasanova2019survey, chen2020survey}, up to our knowledge, this research is the first to focus on design vulnerabilities and map them to their related entries in the Common Weakness Enumeration (CWE) list, which is a catalog of common software and hardware weakness types \cite{CWE}. Additionally, different from previous efforts that performed empirical evaluations of automated analysis tools to compare their real capabilities \cite{durieux2020empirical, 10.5555/3291291.3291303, leid2020testing}, this work  characterise a set of automated analysis tools that can discover a group of design vulnerabilities pertinent to smart contracts.


\section{PRELIMINARIES}
\label{section:preliminaries}

\subsection{Smart Contracts and Ethereum}
A smart contract is stored and run on blockchain technology, a distributed database that stores transactions in a decentralised sequence of blocks. Thus, the correct execution of the contract is enforced by the blockchain consensus protocol \cite{wang2019blockchain}. 

Ethereum is the most popular general-purpose blockchain platform that enables developers to deploy smart contracts written using Turing-complete languages such as Solidity. Solidity is an object-oriented language designed specifically for writing blockchain contracts. The programmable contracts are then compiled into bytecode that executes on the Ethereum Virtual Machine (EVM). EVM executes typical cryptocurrency transactions and contracts bytecode, which are a special kind of transaction \cite{wood2014ethereum}. 

In practice, several properties distinguish smart contracts from regular computer programs and make the implications of deploying vulnerable contract more severe.  First, the primary distinction is that contracts operate on a decentralised public blockchain network that makes the contract open for inspection, with the state of the contract transparent and traceable by everyone \cite{wang2019blockchain}. Second, contracts typically handle monetary transactions that can involve considerable amounts of Ether, an Ethereum cryptocurrency with a current market value equivalent to billions of dollars \cite{chen2020survey}. The combination of monetary value and public availability makes vulnerable smart contacts compelling targets for attackers. Third, smart contracts are irreversible and immutable. Thus, vulnerable contracts cannot be patched after deployment to the blockchain \cite{luu2016making}. The only way to fix the contract, is to deploy a new version with a repair code. However, the old version will remain in the blockchain. Fourth, deploying and executing smart contracts cost an amount of gas (the fuel of computation in Ethereum) \cite{yu2020smart}. 

To prevent the abuse of computational resources, in Ethereum, developers and users are required to pay a gas fee to deploy and to execute contracts, respectively \cite{wood2014ethereum}. The concept of gas involves the following: (i)
\textbf{gas cost}, which is a constant amount that determines the computational effort required to execute specific operations; (ii) \textbf{gas price}, which denotes the amount of Ether that users are required to pay for each unit of gas; it is dynamic, governed by Ethereum miners, and measured in Gwei (1 Gwei = 0.000000001 Ether); and (iii)
\textbf{gas fee}, which is the incentive received by miners in exchange for the computional resources used to execute the operations of a contract and the building of blocks; gas fee is converted to Ether and paid to miners.  
 


\subsection{The Common Weakness Scoring System} \label{CWSS}
The Common Weakness Scoring System (CWSS™) \cite{CWSS_new} provides a quantitative mechanism to score unfixed CWEs found in software. The system supports a prioritisation of weaknesses in terms of their potential negative consequences. The scoring formula is based on three metrics: \textit{Base Finding}, which consists of several factors that estimate the level of technical impact once the weakness is successfully exploited, the accuracy of finding, and the effectiveness of the existing control; \textit{Attack Surface}, whose factors estimate how easy the attacker could exploit the weakness;  \textit{Environmental}, which involves factors that refer to a specific operational context such as the potential impact to the business and the likelihood of discovery. Each metric factor is assigned a numeric value that is calculated based on specific formulas, and the sub-score of each metric is multiplied with each other to generate the final CWSS score in a range between 0 and 100. 
 
\subsection{Technical Debt}
Technical debt is a metaphor devised to capture how the value of software engineering decisions evolves over time \cite{debtexchanges}. Specifically, the metaphor supports the identification of the roots of a sub-optimal decision, the estimation of its value, and the monitoring of the environmental trade-offs in which the decision was made \cite{kruchten2012technical}. The metaphor also supports attributes that are intrinsic to debts in finance such as principal and interest \cite{tom2013exploration}. In the metaphor, the principal represents the value of the gap between the ideal and the actual decision, whereas the interest represents the additional effort that needs to be incurred to pay back the principal. 

Technical debt has been applied in architectural design to value the gap between the ideal and an actual  decision, to identify the architectural root of an issue, and to manage the debt incurred by a decision \cite{tom2013exploration}, among others. Since some roots of security issues tend to be intertwined with architectural decisions, the metaphor has also been used to identify security vulnerabilities \cite{nord2016can}, prioritise the attention to security weaknesses \cite{izurieta2019leveraging}, and  to estimate the consequences on business value if vulnerabilities are exploited \cite{izurieta2018position}.

\section{Our Approach for Assessing Technical Debts} 
\label{section:approach}
This section presents our approach to assessing the security technical debts incurred in smart contracts design. The assessment involves detecting design issues and quantifying their consequences to security if remain unfixed.

We have defined the steps that support the creation of secure by design smart contracts: 

\begin{enumerate}
    \item Identify security design vulnerabilities in a smart contract. 
    \begin{enumerate}
        \item Run automated security analysis tools on the smart contract source code to obtain a list of potential vulnerabilities.  
        \item Perform a manual analysis complementing the automated analysis to uncover any missed vulnerability. 
        \item Determine the design vulnerabilities and map them to related weaknesses to highlight the root cause of the issues. 
        \item Classify the design vulnerabilities per design flaw categories to determine the negative technical impact upon the contract.
    \end{enumerate}
    \item Measure the ramifications of the identified design vulnerabilities.
     \begin{enumerate}
     \item Estimate the monetary cost of technical debt principal by calculating the gas fee for redeploying a patched version of the vulnerable contract. 
     \item For each vulnerability identified, estimate technical debt interest value by quantifying the negative security consequences and the interest growth rate over time.
     \end{enumerate}
\end{enumerate}
Our approach is a debt-aware assessment approach, as it facilitates visualisation of the debt incurred by exploitable flaws created while designing smart contracts. Furthermore, applying this approach, developers can be aware of the long-term consequences of security design issues. Subsequently, the developer can prioritise the debt based on both the monetary cost and the interest value related with vulnerability violations.

\subsection{Identification of Security Design Vulnerabilities} \label{subsection 3.1} 
Building our assessment approach involves two steps: aggregating vulnerabilities caused by flaws in smart contracts design, Subsection \ref{subsection 3.1.1}, and selecting security analysis tools that assists the identification process, Subsection \ref{subsection 3.1.2}.    

\subsubsection{Mapping Design vulnerabilities to Security Design Weaknesses} \label{subsection 3.1.1} 
Security flaws coming from design decisions have been reported as the main cause of software security problems \cite{nord2016can}. If these flaws remain unaddressed, these can be viewed as root cause of technical debt, with consequences observed through accumulation of interests over time \cite{izurieta2018position} \cite{nord2016can}. Additionally, if security software engineers are aware of these issues but they do not fix them, these workarounds can be considered as (self-) admitted technical debts.

Although several efforts, from industry and academia, illustrate smart contract vulnerabilities, they have missed to distinguish between coding and architectural design vulnerabilities. Therefore, in this study, we present a set of vulnerabilities rooted in the architectural design of smart contracts and their related security weaknesses. The aggregated set is classified based on the security impact of such issues.

Security weaknesses are flaws in software that may lead to exploitable security vulnerabilities. Therefore, the identification of weaknesses can assist us in understanding the security problems and performing root cause analysis. Our study leverages Common Weakness Enumeration (CWE™) \cite{CWE}, which is a community-developed list of common security weaknesses that may appear in architecture, design, or implementation of software.
Thus, we aim to map vulnerabilities in contract design to security weaknesses in the CWE catalog that stands out regarding adoption and scope of coverage. Although CWE does not refer to any weaknesses particular to smart contracts, it depicts associated weaknesses at higher abstraction layers. 

We followed a systematic procedure to collect and map each design vulnerability in contract design architecture to the corresponding weaknesses in CWE. This procedure consisted of two steps. 

First, we collected a comprehensive list of security vulnerabilities from academic papers that surveyed existent vulnerabilities in smart contracts \cite{hasanova2019survey} \cite{li2020survey} \cite{chen2020survey} \cite{atzei2016survey}; we also considered Ethereum community \cite{openzeppelinforum}, wiki \cite{githubwiki}, and developers' blogs \cite{mediumblog} \cite{YosRiady} that listed and explained exploitable flaws  not discussed in the literature. Specifically, we extracted information about the vulnerabilities, such as their descriptions, ways of exploitation, and preventive techniques that can be used to avoid them. The collected information assisted in knowing the negative impacts of these issues on the contracts and the cost of fixing them. In addition, the information also helped to determine the vulnerabilities that result from the flaws manifested at the contract's design stages. Second, we analysed each collected vulnerability to map it with the subset of weaknesses (438/1248) of CWE. This subset represents weaknesses that can be introduced during a design stage.




To reduce inherently biases in the mapping process, two authors separately worked over all the collected vulnerabilities. After completing the analysis, results were compared and double-checked with Smart Contract Weakness Classification (SWC) \cite{SWC} Registry. This registry established by a group of developers, auditors, and researchers at ConsenSys Diligence \cite{mythx} that provides smart contracts developers with a system analog to CWE. However, at the time of writing, only 20 architectural design weaknesses are listed in the registry. Finally, each disagreement was discussed with a third author.


We observed that the identified design vulnerabilities can be classified based on their impact into ten categories. Front-Running, Time Manipulation, Denial of Services (DoS), Broken Access Control, Arithmetic Issues, and Bad Randomness. These first six categories are also presented in Decentralised Application Security Project (DASP) \cite{DASP} taxonomy. Other categories are Sensitive Data Exposure and Using Components with Known Vulnerabilities, which are the third and ninth category, respectively in Open Web Application Security Project (OWSAP)-Top 10 Security Risks \cite{owasp}. The two remaining categories are Improper Inheritance and Modularity Violation, which are the flaws that violate object-oriented design principles. Since Solidity is an objected oriented language, the contracts written by this language are also prone to these types of design flaws. 

Table \ref{table:indicators} describes each design flaws category, whereas Table \ref{table:mapping} shows an example of mapping the design vulnerabilities classified under DoS to relevant CWEs. Our replication package, described in Subsection \ref{RP} carries full details of the mapping. 

\begin{table*}[t]
\centering
\footnotesize
\caption{Description of Design Flaws Categories}
\begin{tabular}{@{}llp{1.35\columnwidth}@{}}
\toprule
\multicolumn{1}{c}{\textbf{Category}}                                                                              & \multicolumn{1}{c}{\textbf{Source of Category}}                                                      & \multicolumn{1}{c}{\textbf{Description}}                                                                                                                                                                                                \\ \midrule
Front-Running                                                                         & DASP                                                                    & A type of race condition where a malicious user can steal the solution and submit a transaction with a higher gas price to make their transaction assigned to the block before the victim. \\
Time Manipulation                                                                     & DASP                                                                    & The timestamp of the block is adjusted  by malicious miners to their own advantage.                                                                                                                                            \\
Denial of Services                                                                    & DASP                                                                    & The attacker can make the contract inoperable temporarily or permanently.                                                                                                                                \\
Arithmetic Issues                                                                     & DASP                                                                    & An arithmetic operation that reaches the max or min size of a type and presents incorrect results that compromise contract security and reliability.                                                      \\
Bad Randomness                                                                        & DASP                                                                    & The random number generator is written in a way that is predictable and exploitable.                                                                                                                    \\
Sensitive Data Exposure                                                               & OWSAP-10                                                                & The developer does not adequately protect critical information related to the contract and assumes that private type variables cannot be read.                                                                    \\
\begin{tabular}[c]{@{}l@{}}Using Components with\\ Known Vulnerabilities\end{tabular} & OWSAP-10                                                                & Malicious or deficient components, such as libraries or off-chain data sources, where the developer unaware of all the running code.                                               \\
Improper Inheritance                                                                  & \begin{tabular}[c]{@{}l@{}}Object Oriented \\ Design Flaws\end{tabular} & When a contract inherits another contract, the presence of multiple variables with the same name in both contracts might lead to unintended effects.                                                                                                 \\
Modularity Violation                                                                  & \begin{tabular}[c]{@{}l@{}}Object Oriented\\ Design Flaws\end{tabular}  & This consists of a tight coupling between contracts that makes them frequently change together.                                                                                                                                \\ \bottomrule
\end{tabular}
\label{table:indicators}
\end{table*}

Our classification allows the designers to select a specific design flaws category and visualise all related vulnerabilities and weaknesses. It may also serve as a guide for architects to avoid common security architectural issues when creating smart contracts.


\subsubsection{Selection of Security Analysis Tools} \label{subsection 3.1.2}

The analysis of potential vulnerabilities in smart contracts needs to be performed before its deployment to the immutable environment of the blockchain, in which refactoring or updating the contract is costly and not trivial.


Recently, a growing number of security analysis tools have emerged to detect common vulnerabilities and bad practices in Ethereum smart contracts written in Solidity. Although these tools have been developed by different teams such as academic teams \cite{10.1145/3243734.3243780}, community teams \cite{luu2016making}, and industrial teams \cite{manticore}, most of the existing tools are inaccurate in finding the security issues as they yield a considerable number of false positives \cite{durieux2020empirical} \cite{10.5555/3291291.3291303}. Furthermore, each tool only detects a limited scope of vulnerabilities. Consequently, the combination of several existing tools is essential to increase coverage and accuracy of vulnerabilities detection.

\begin{table}[b]
\caption{A Vulnerabilities Mapping for DoS to CWE Categorisation}
\begin{tabular}{@{}p{0.35\columnwidth}p{0.15\columnwidth}p{0.40\columnwidth}@{}}
\toprule
Design Vulnerabilities           & CWE     & Design Weaknesses                                                                                    \\ \midrule
Exception handling problem       & CWE-703 & \begin{tabular}[c]{@{}l@{}}Improper Check or Handling\\ of Exceptional Conditions\end{tabular}  \\
Non-validated arguments          & CWE-20  & \begin{tabular}[c]{@{}l@{}}Improper Input Validation\end{tabular}                                 \\
DoS by external contract/Call    & CWE-703 & \begin{tabular}[c]{@{}l@{}}Improper Check or Handling \\ of Exceptional Conditions\end{tabular} \\
Calculates the upper bond of Gas & CWE-400 & \begin{tabular}[c]{@{}l@{}}Uncontrolled Resource \\ Consumption\end{tabular}                         \\
Costly pattern/Costly loop       & CWE-400 & \begin{tabular}[c]{@{}l@{}}Uncontrolled Resource \\ Consumption\end{tabular}                         \\
Reachable SELFDESTRUCT           & CWE-28  & \begin{tabular}[c]{@{}l@{}}Improper Access Control\end{tabular}                                   \\ \bottomrule
\end{tabular}
 \label{table:mapping}
\end{table}

In this study, we selected a set of security analysis tools that help in identifying design vulnerabilities in smart contracts. To do that, we investigated the academic literature, searched the Internet, and scanned Github.  
The tools selected for our study were those matching the following criteria (C): \\\textbf{C\#1.} The tool is free, publicly available and specific information about it can be found in at least one official source.\\ 
\textbf{C\#2.} The tool is up to date.\\ 
\textbf{C\#3.} The tool can take source code of the contract as an input.\\ 
\textbf{C\#4.} The tool identified at least one design vulnerability or bad practice related to design. 

After exploring all the collected tools, we identified only 9 tools that met the criteria. These tools are: Slither \cite{8823898}, SmartCheck \cite{tikhomirov2018smartcheck}, and Securify \cite{10.1145/3243734.3243780} which perform static analysis techniques. Mythril \cite{mueller2018smashing} which is a symbolic analysis tool and Manticore \cite{manticore} which perform dynamic symbolic execution identification. Another tool is sFuzz \cite{10.1145/3377811.3380334} which applies a fuzzing testing technique. Solhint \cite{solhint} and Ethlint \cite{Ethlint} perform static analysis to identify security issues and bad practices. Final tool is Mythos which applies a combination of dynamic symbolic execution and fuzzing techniques. 


\subsection{Measurement of Negative Consequences of Design vulnerability in Smart Contracts} \label{subsection 3.2}
Besides detecting vulnerabilities, it is also crucial to estimate the associated implications of taking a shortcut by not fixing those issues. To this end, we adopt the notions of principal and interest to quantify the debt cost and value related to each identified vulnerability in a smart contract. Quantifying the cost and value of the security debts aids developers to apply a cost-effective technique and justifies the investment in fixing security problems. 

\subsubsection{Estimation of Gas Cost (Quantifying the Principal)} \label{subsection 3.2.2}
Immutability in smart contracts restricts developers’ ability to patch vulnerabilities after deploying the contract to the blockchain. The developer needs to upload a new version of the contract after fixing those vulnerabilities, and this requires additional gas consumption with the associated expenses. As a result, the need to refactor vulnerable deployed contract has financial implications.

Our aim is to calculate the gas consumption fee for redeploying a contract to estimate technical debt principal (P). This estimation provides useful insights for developers regarding the cost of repairing vulnerabilities in the deployed contracts. We use the following formula:
\setlength{\abovedisplayskip}{3pt}
\setlength{\belowdisplayskip}{3pt}
\begin{equation} \label{E1}
P_{Security Debt} = Gas\_D(c) + Gas\_U(c) 
\end{equation} 
where Gas\_D indicates the cost of gas required to deploy the repaired contract (c), and Gas\_U indicates the cost of the gas required to update the contract (c) for a given issue. 

\textbf{Gas Cost Required to Redeploy Repaired Contract.} 
The cost of the gas required for deployment depends on the size of the smart contract. The number of functions (NoF) and lines of code (LoC) are both influencing factors, and the more complex the contract is, the more gas consumption is required.  The uploading cost fee is computed using {\itshape gas\_price × gas\_cost}, where the former is the value of a unit of gas as specified by the market, and the latter is mostly determined by the summation of the following factors: \\ 
$G_{create}$, 32000 gas,      paid for a CREATE operation, \\
$G_{transaction}$, 21000 gas, paid for every transaction,  \\
$G_{codedeposit}$,  200 × |o| gas, paid per byte for a create operation, |o| amount of bytecode in the compiled contract, \\
{\itshape Execution costs}, gas paid for necessary computation processes. 

The last factor refers to the costs associated with the part of the code that requires to be executed before the creation of the contract, such as initialising the state variables whose values are permanently stored in the contract storage, and executing a constructor. If the constructor requires a lot of computation to generate the bytecode, then there will be extra expense. Appendix G in the Ethereum yellow paper \cite{wood2014ethereum} shows the costs, in gas, of several opcode operations.

\textbf{Gas Cost of Update Pattern.} Since redeployment results in a new contract with a new address, an update pattern needs to be utilised to avoid using the vulnerable deprecated contract. The developer can use a self-destruct opcode, which costs $Gselfdestruct = 5000$ gas, to destroy the contract. Another option is to upload the main contract with a proxy smart contract, which has a changeable variable that stores the main contract's address. Thus, once an updated contract version is released, the value of this version is updated.

\subsubsection{Estimation of Security Consequences (Quantifying the Interest)} \label{subsection 3.2.1}
In the security context, the interest value represents the undesirable effects that can result if the vulnerability is exploited. The longer the exploitable design flaw remains unaddressed in the deployed contract, the higher the chance the debt interest associated with this flaw grows. Accordingly, three factors are considered when estimating the accumulated interest: CWSS score, contract activity level, and contract lifespan.

\textbf{CWSS Score.} We adopted the CWSS to estimate the severity of identified weaknesses in a contract. The CWSS framework provides different scoring methods; in this study, we use the targeted method, which assesses individual design weakness. Multiple factors are used to quantify CWSS scores in smart contracts, such as mapping the vulnerabilities to the related CWEs and the proposed design flaws categories. This information allows the negative impact on the contract in the case of an attack to be estimated. The estimated score generated by CWSS allows for technical debt items to be visualised and those debts that lead to more severe consequences to be addressed. In practice, addressing contract security design issues early, in the pre-deployment stage, minimises debt.

\textit{Accumulated interest.}
Debt interest increases when exploited contracts become extremely common like, for example, the hacked Parity Multi-Sig Wallet contract \cite{chen2020survey}. This was a smart contract for a multiple signature wallet that had a critical, exploitable vulnerability, which allowed an attacker to steal millions of dollars. The debt interest also increases if the exploitation of vulnerabilities in the contract has irreparable consequences, as is the case with suicidal vulnerability. With this vulnerability, any arbitrary account can kill a contract, causing it to stop functioning and locking its ether \cite{10.1145/3243734.3243780}.  
The accumulation of interest associated with vulnerabilities in smart contracts is difficult to quantify; however, we find that the activity level and the lifespan of the contract are useful for estimating interest growth.

\textbf{Contract Activity Level (CAL).} CAL refers to the expected number of active users and the number of transactions that a contract is expected to deal with. This factor can be predicted from statistical information about contracts provided on websites, such as State of the ÐApps \cite{StateoftheDApps}, which is a website with a curated list of smart contract-based applications. This website ranks the contracts and categorises them based on their activity level. Statistical research  \cite{oliva2020exploratory} shows that the top three high-activity contract categories in 2020 are games, currency exchanges, and gambling. 

\textbf{Contract Lifespan (CLS).} CLS refers to the time, measured in days, between the contract’s deployment and its last execution. Most smart contracts live between 2 and 800 days \cite{lohr2020maintenance}, and the average age of smart contracts is 295.6 days \cite{oliva2020exploratory}. A contract can be short-lived, medium-lived, or long-lived \cite{oliva2020exploratory, lohr2020maintenance}. A contract is long-lived when it is expected to be executed for a long interval or even for as long as the network exists. Conversely, a short-lived contract is mostly designed to be executed for a limited time and then either the interaction with it ends or it is destroyed. This type of contract typically includes few operations, involves few fixed users, and has limited transactions. Thus, the security implications of breaching this contract are not significant compared to a long-lived, highly active contract in which there is interaction with thousands of customers and thousands of transactions are accepted over a long period of time. 

Therefore, during the CLS period (i.e., before the contract reaches its maturity stage), the accumulated interest (AI) of the security technical debt can be estimated as follows: 
\setlength{\abovedisplayskip}{3pt}
\setlength{\belowdisplayskip}{3pt}
\begin{equation} \label{E2}
AI_{Security Debt} = CWSS score \times CAL \times CLS
\end{equation} 
Nevertheless, we acknowledge that some issues may go beyond, and the interest can be adjusted accordingly. For this paper, we focus on interest within the CLS period, as this enables action to be taken before contract redemption to avoid the accumulation of the issues that are rooted in technical debt. 

The developer can assign a point scale to each contract activity level category and lifespan category. The multiplication of the scores of the three factors determines the accumulated interest.

\subsection{Replication Package for Replicability} \label{RP}
The data source and replication package for our experiments are publicly available \footnote{https://bitbucket.org/Smart\_Contract/assessing-smart-contracts-security} to assist in verifiability and reproducibility of our results.The package includes: (i) the complete mapping of all aggregated design vulnerabilities and their rated CWE entries; (ii) the list of all founded analysis tools; (iii) the set of vulnerabilities claimed to be identified by each tool; (iv) details of inclusion criteria that were not met by each of the excluded tools; (v) the dataset used in the experiment; and (vi) the detailed results of each step of the approach. 

\section{Experimentation}
\label{section:experimentation}
To demonstrate our approach, the following example shows how the steps of our approach are performed to assess the potential security-related debts in smart contracts.
\subsection{Experiment Setup}
We collected a dataset of 16 representative vulnerable smart contracts that are either actual contracts  identified as vulnerable or explicitly programmed to demonstrate a specific vulnerability. This dataset comes from several publicly-available resources: (i) \textit{GitHub} repositories such as SWC Registry, not-so-smart-contracts \cite{not-so-smart-contracts}, and Solidity-Security \cite{solidity-security-blog}; (ii) \textit{Ethernet} \cite{ethernaut}, an online game for smart contract attacking challenges; and (iii) \textit{blog.positive} \cite{blog.positive}, which publishes articles discussing vulnerabilities in smart contracts. 

We intentionally chose a dataset of known vulnerable contracts to examine the ability of the 9 selected security analysis tools to identify vulnerabilities. Our selection of smart contracts enables a discussion of their source codes and permits the publication of our results for replication without users being involved in legal and privacy issues.

All the tools are installed and run on the same computer utilising Solidity compiler- solc (v 0.7.1), under Ubuntu 18.04.5 operating system. Only sFuzz provides a web-based interface for using the tool, then there was no need to install anything related to it. 
\begin{table}[t]
\centering
\footnotesize
\caption {Experiment Dataset. For each vulnerable contract, we provide design flaws categories of its flaws (DFC), number of design vulnerabilities (Vulns), and lines of code (LOC).  }
\begin{tabular}{@{}llcc@{}}
\toprule
\multicolumn{1}{c}{\textbf{Contracts}} & \multicolumn{1}{c}{\textbf{DFC}}                                             & \textbf{\#Vulns} & \textbf{\#LOC} \\ \midrule
FindThisHash                           & Front-Running                                                                & 1                & 9              \\
EtherLotto                             & \begin{tabular}[c]{@{}l@{}}Time Manipulation /\\ Bad Randomness\end{tabular} & 1                & 20             \\
Roulette                               & Time Manipulation                                                            & 1                & 14             \\
Lottopollo                             & \begin{tabular}[c]{@{}l@{}}Time Manipulation /\\ Bad Randomness\end{tabular} & 1                & 24             \\
DosAuction                             & DoS                                                                          & 1                & 13             \\
SimpleToken                            & \begin{tabular}[c]{@{}l@{}}DoS /\\ Access Control Broken \end{tabular}       & 1                & 65             \\
Etheraffle                             & DoS/ Bad Randomness                                                          & 2                & 122            \\
DosNumber                              & DoS                                                                          & 1                & 28             \\
AccessControl                          & Access Control Broken                                                        & 1                & 53             \\
BlockdBuildDemo                        & Access Control Broken                                                        & 3                & 62             \\
FunctionTypes                          & Access Control Broken                                                        & 2                & 18             \\
OddEven                                & Sensitive Data Exposure                                                      & 1                & 22             \\
Transaction\_malleablity               & Sensitive Data Exposure                                                      & 1                & 77             \\
Token                                  & Arithmetic Issues                                                            & 1                & 17             \\
TokenSaleChallenge                     & Arithmetic Issues                                                            & 1                & 20             \\
CEOThrone                              & Improper Inheritance                                                         & 1                & 21             \\ \bottomrule
\end{tabular}
\label{table:dataset}
\end{table}

\subsection{Experimental Study}
\begin{figure}[t]
\centering
\includegraphics[width=0.49\textwidth]{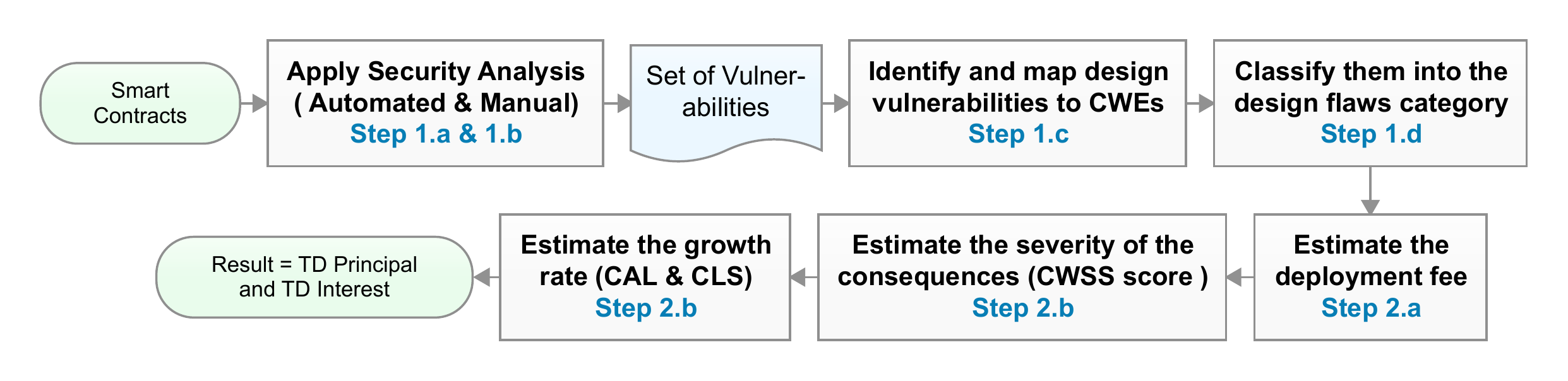}
\caption{Experiment execution steps}
\label{F1}
\end{figure}

Figure \ref{F1} represents the overall procedure of our experiment.

\textbf{Step 1.a.} We first ran the automatic security tools against each contract in our dataset to identify the design vulnerabilities. As mentioned in the previous section, different types of tools were selected to enhance the detection abilities of design vulnerabilities. We started the operation process by running the tools that perform static analysis techniques. After the static analysis has been completed, operating the tools that perform dynamic analysis techniques was the next step for obtaining deeper results. We finalised the analysis by running the tool’s applied fuzzing techniques. 

\textbf{Step 1.b.} We conducted a manual analysis to complement the previous step as the 9 tools could not detect all the known vulnerabilities in the dataset. We annotated the known vulnerabilities and detected several other flaws that were not mentioned in the online source with regard to the vulnerable contracts. We ended up with a set of possible vulnerabilities that was composed of flaws not related to the design. 

\textbf{Step 1.c.\&1.d.} We checked the aggregated set of design vulnerabilities and the aligned CWEs to only record the issues that belonged to that set. Each contract had one or more vulnerabilities mapped in at least one of the design flaw categories. There were 20 distinct vulnerabilities in our dataset. Table \ref{table:dataset} shows each collected contract, its name, the categories of its flaws, the number of design vulnerabilities, and the lines of code. 

\textbf{Step 2.a.} The cost fees required to deploy the patches contracts were calculated by Formula \ref{E1} to quantify the technical debt principal. A self-destruct updated pattern were considered when calculating the total gas cost. At the time of experimenting, the average gas price was 126 Gwei, and Ether's price was around \$500 US. 

\textbf{Step 2.b.} We estimated the security risk severity scores related to identified vulnerabilities and weaknesses using the CWSS formula that explained in Subsection \ref{CWSS} . 
We used the OWSAP method to estimate the technical and business impact factors, in case vulnerabilities are exploited. This method suggests dividing the technical impact into sub-factors aligned with the traditional security areas of concern: loss of confidentiality, integrity, availability, and accountability. Additionally, our proposed categories of design flaws support determining the technical impact as each category shows the ramifications of the exploitation.
Similarly, the method divides the business impact factor into sub-factors common to many businesses: financial damage, reputation damage, non-compliance, and privacy violation. Each sub-factor has several scored options. We selected one of the options and then averaged the scores for each factor, technical and business impacts, to decide their severity level. 

There are other required factors to calculate the CWSS score, such as required privileges; authentication and access vector to perform the attack; and the acquired privilege after accomplish it. These factors were determined based on the information collected about each vulnerability and the way of exploiting it. The knowledge about the vulnerable contracts and security analysis performed assisted in determining factors such as finding confidence, likelihood of discovery and exploit. 

In addition to the CWSS score, the contract’s activity level and lifespan factors have to be estimated to visualise the potential accumulated interest. A six-point scale is used to determine the activity level, where 6 refers to the top 3 highly active contract’s category, and 1 refers to the lowest 3 categories. The ranking of the categories is informed by the State of the DApps website. Since most smart contracts live between 2 to 800 days, we divided the lifespan into three intervals and gave each interval a score as follows: (i) Short-lived, 1-266 days, 0.17 score; (ii) Medium-lived, 267- 533 days, 0.35 score; (iii) Long-lived, 534-800+ days, 0.5 score. 
The accumulated interest for each unfixed vulnerability were calculated by Formula \ref{E2}. The value of the final score is between 0 and 300. Noticeably, we set those particular scores to visualise how the interest could redouble. However, contract developer can customise the scores based on their context. 

\section{Results and Discussion}
\label{section:discussion}
In this section, the collected dataset of some represented vulnerable contracts are analysed and discussed with respect to our research questions. Additionally, we discuss the practical implications of our findings and we outline the potential threats to their validity.

\subsection{Identification of Design Vulnerabilities (RQ1)}
\begin{figure}[t] 
\centering
\includegraphics[width=0.45\textwidth]{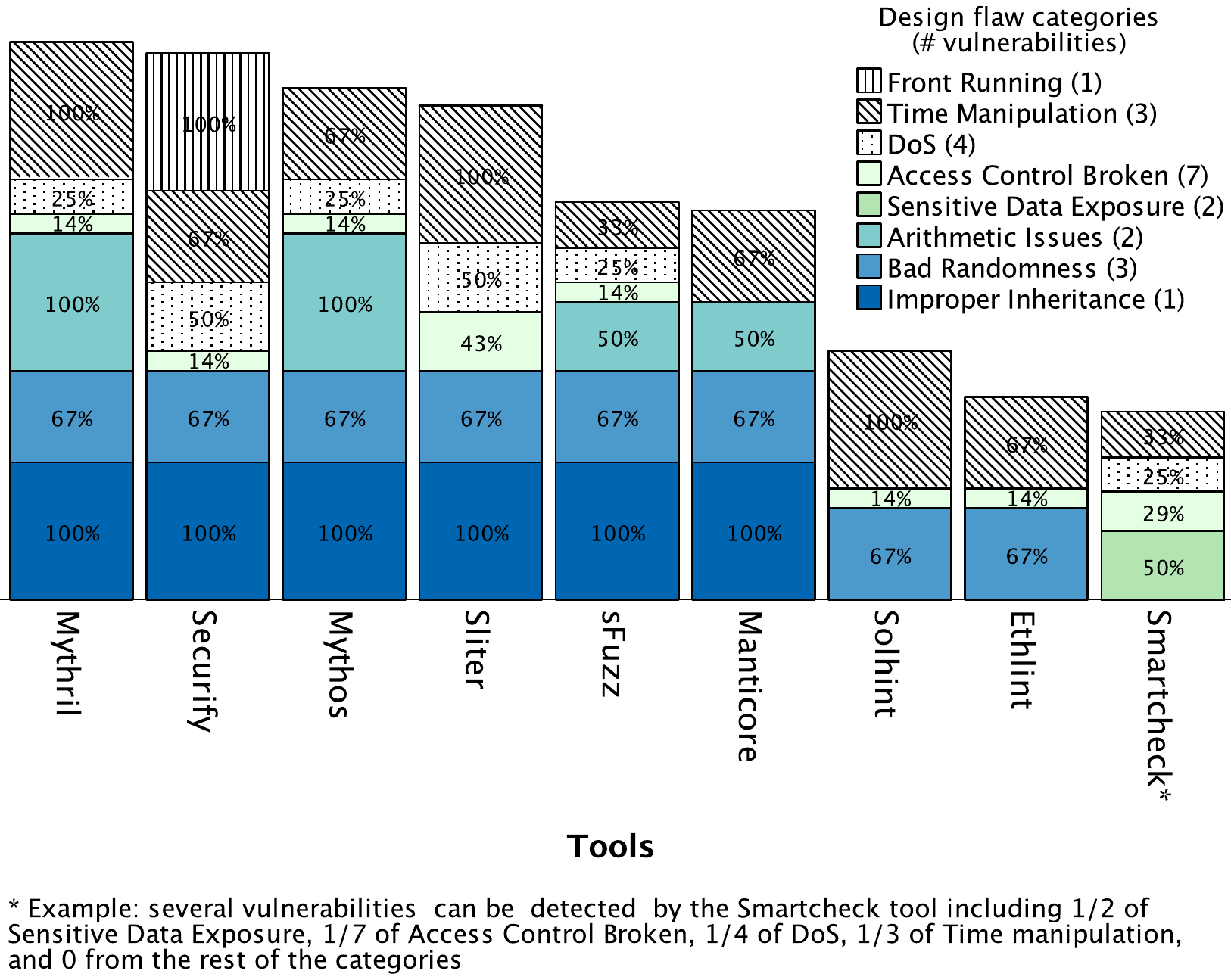}
\caption{Identified vulnerabilities per category by each tool.}
\label{F2}
\end{figure}

Locating design vulnerabilities in smart contracts source code is the main step in our assessment approach. Figure \ref{F2} presents the results of executing 9 security analysis tools on 16 vulnerable contracts in order to locate their design vulnerabilities. It shows the percentage of vulnerabilities that each tool was able to detect per design flaws category. 

Nine tools, in combination, were only able to locate 70\% (14/20) of all the vulnerabilities. Most of the tools were generally able to detect a high percentage of vulnerabilities classified in the categories \textit{Time Manipulation}, \textit{Improper Inheritance}, and \textit{Bad Randomness}. However, the tools under performed when it came to identifying vulnerabilities of the categories \textit{Front-Running} and \textit{Sensitive Data Exposure}. The 9 tools failed to detect four vulnerabilities linked to \textit{Access Control Broken}, one vulnerability linked to \textit{DoS}, and another linked to \textit{Sensitive Data Exposure}. Throughout the analysis, we observed that the output of some tools was noisy, as they failed to disregard the false warnings and detected unreal flaws.

Figure 2 indicates that the tools demonstrate different ability to locate design vulnerabilities. The tool Mythril surpassed all other tools as it was able to identify more vulnerabilities in the 16 contracts than any of the others. Securify is the only tool that detected the vulnerability, Transaction Ordering Dependency, linked to the \textit{Front-Running category}. Although the documentation of several tools such as Manticore claimed that they could detect this vulnerability, they failed to identify it. Similarly, Smartcheck is the only tool that identified the vulnerability, Unencrypted Private Data On-Chain, in the \textit{Sensitive Data Exposure} category. However, the output of Smartcheck does not reveal a clear explanation of the identified flaws. Mythril and Mythos both provide informative output as they link the identified vulnerabilities to related SWC. Securify resulted in the most confusing output among all the tools because it showed a large number of false alarms.

Our study emphasises that conducting manual analysis to complement automated ones is essential to discover most vulnerabilities. This is because, as our analysis experiment shows, the scope of issues addressed by state-of-the-art tools is limited. Additionally, the accuracy of their analysis needs to be increased by reducing the likelihood of false-positive alarms that confuse the developers and discourage them from leveraging the results. Even though an empirical study \cite{durieux2020empirical} claimed that combining different types of analysis tools yields more vulnerability coverage, the existing tools are not powerful enough to replace manual analysis. Thus, taking a shortcut by deploying the contract to the public without performing both automated and manual inspection processes might lead to invisible exploitable flaws that are prone to attacks. 

Through the analysis, all the detected flaws were found in our aggregated set of design vulnerabilities and their associated weaknesses. In contrast, some of these detected flaws, including no restricted write and no restricted transfer, did not exist in the SWC Registry which provides a set of classified issues that come up in smart contract development. Moreover, our proposed design flaw categories cover all the detected flaws, while the current DASP taxonomy is not extensive enough to include all types of security design issues that affect smart contracts. Unlike DASP, our classification includes vulnerabilities of the following categories: Sensitive Data Exposure, Using Components with Known Vulnerabilities, Improper Inheritance, and Modularity Violation. 

Classifying vulnerabilities based on their impact and mapping them to a wider group of agreed-upon weaknesses (CWEs) come with several benefits: (i) it provides a common language for defining security architectural design issues in smart contracts; (ii)  it offers a simple way to classify security issues in such contracts; and (iii) it generates useful insights into the root-causes of vulnerabilities, and the negative impacts posed by the exploitable ones.
\begin{figure}[t]
\centering
\includegraphics[width=0.47\textwidth]{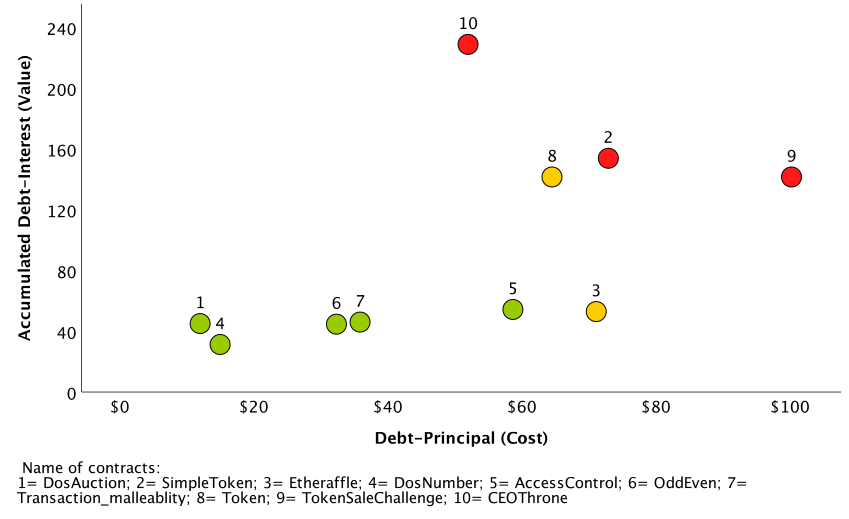}
\caption{Principal and interest of ten vulnerable contracts.}
\label{F3}
\end{figure}
\subsection{Estimation of Negative Consequences (RQ2)}
The second step in our approach is to estimate the ramification of deploying vulnerable contracts to the public blockchain in terms of debt principal and accumulated interest. Figure \ref{F3} shows debt principal and interest of 10 vulnerable contracts from our dataset out of 16. The figure represents the estimated monetary cost of refactoring each contract and the value of the accumulated interest in relation to the effects of the vulnerabilities over time, if they are left unfixed. 

For the sake of visualising and discussing the data, we have categorised the coordinates of the (cost, value) points in the graph into three categories: high, medium, and low, depending on their severity. Given the score for values in this range [0,300], we defined (100<value<=200) as medium, values greater than 200 as high, and values lower than or equal 100 as low. As in our example the highest cost is \$100.25, we defined (\$33.33<cost<=\$66.66) as medium, a cost greater than \$66.66 as high, and a cost lower than or equal to \$33.33 as low. The final degrees of severity are then derived from both the cost and the value using Table \ref{T4}. As such, vulnerabilities in contracts 2, 9, and 10 fall into the high category; vulnerabilities in contracts 3 and 8 into the medium category; and those in contracts 1, 4, 5, 6, and 7 into the low category.

Estimating the cost and value related to security violations in smart contracts facilitates decision-making regarding which issues need more attention when seeking to reduce technical debt. For instance, a vulnerability in the CEOThrone contract is assumed to be in the high category. The cost of gas required to fix the issue that has not been fixed before deployment is 897,200, which is equivalent to \$51 at the time of writing. The size of this contract is relatively small compared to other contracts in our dataset. There is only one operation in the constructor, and there are no state variables that need to be initialised. 

However, the contract suffers from variable shadowing vulnerability, which occurs when a variable declared within a child contract has the same name as a variable declared in a parent contract. In the case of CEOThrone, this allows an unauthorised node to withdraw the balance of the contract, leading to a significant technical and business impact. The likelihood of discovering and exploiting this issue is also high, since most of the analysis tools were able to detect it. The vulnerability in this contract allows any attacker to gain the required privilege to exploit it. CEOThrone is a contract created for a game and it can thus be assumed that the contract will be highly active and have a long lifespan. As such, we estimated the accumulated security debt interest for this contract at around 228.9, which is a high score compared with other contracts. 

Beyond the results of our experiments, our assessment approach can significantly reduce the cost of developing smart contracts, in particular in cases where a contract is big and, as such, has a potentially higher cost, or in the case of fixing an issue in a contract that requires modifying and redeploying other dependent contracts. We provided a quantitative way of indicating the relative costs and values of publicly deploying vulnerable contracts. Developers can thus make informative decisions before uploading a contract to the blockchain. Additionally, they can prioritise the resolution of issues based on the available quantitative information. Overall, the assessment approach presented here supports reducing the introduction of unintentional debt caused by unawareness of security issues rooted in smart contract design while increasing the visibility of such issues and helping manage the debt more strategically. 

\subsection{Threats to Validity}
A potential threat to internal validity is related to the possibility of considering alternative security flaws scoring methods. However, unlike other scoring systems, the CWSS can be applied at the early stages of the development process. CWSS provides support if there is incomplete information, as most of its factors have values for uncertainty and flexibility, such as Unknown and Not applicable. Furthermore, CWSS facilitates a thorough estimation of the security consequences of uploading vulnerable contracts. It considers 16 factors compared to the only four factors in the risk rating of the OWSAP method.

Another threat relates to the potential subjectivity of CWSS results. We mitigated this possible risk as follows: (1) we proposed categories of design flaws that enable precise determination of technical impact; (2) conducting the security analysis creates awareness of the ease and likelihood of discovering any flaws; (3) in the proposed estimation Formula \ref{E2}, the CWSS score is accompanied with smart contract activity level and lifespan factors. 

A potential threat to external validity relates to the fact that the contracts we have considered in our experiment are relatively small and simple (approximately 1.5KB) - typical size of commonly used contracts in practice as contracts are recommended to be simple and not complex. Nevertheless, the same steps and analysis can scale up to larger ones, often not exceeding 24KB in practice. Referring to our 10 out of 16 selected contracts in our study, the estimated monetary costs required to pay for repairing them were not significant. However, the quantitative approach provided for calculating the debt principal is applicable to any size and type of smart contract. As mentioned earlier, the monetary cost is dynamic as it depends on the gas price and the Ether price, either of which may significantly increase. For instance, in January 2018 the price of Ether rose dramatically to \$1066 US. Hence, uploading even a small and simple contract might cost a significant amount of money.
\begin{table}[t]
\centering
\footnotesize
\caption{Overall severity level}
\label{T4}
\begin{tabular}{c|c|c|c|}
\cline{2-4}
                                      & \multicolumn{3}{c|}{\textbf{Value}} \\ \hline
\multicolumn{1}{|l|}{\textbf{Cost}}                      & \cellcolor{mygray}\textbf{High}         & \cellcolor{mygray}\textbf{Medium} & \cellcolor{mygray}\textbf{Low}   \\ \hline
\multicolumn{1}{|l|}{\cellcolor{mygray}\textbf{High}}    & \cellcolor{orange}Critical          &    \cellcolor{red}High        &  \cellcolor{yellow}Medium    \\ \hline
\multicolumn{1}{|l|}{\cellcolor{mygray}\textbf{Medium}}  & \cellcolor{red}High          &    \cellcolor{yellow}Medium   &  \cellcolor{green}Low    \\ \hline
\multicolumn{1}{|l|}{{\cellcolor{mygray}\textbf{Low} }}  &  \cellcolor{yellow} Medium   &   \cellcolor{green}Low        &  \cellcolor{green}Low     \\ \hline
\end{tabular}
\end{table}

A potential threat to construct validity is the completeness of the aggregated set of design vulnerabilities. We mitigated this risk by conducting a thorough inspection of the academic papers, the Ethereum community, Wiki pages, and developers’ blogs. Nevertheless, the set of vulnerabilities supplied is continually evolving because of the high potential for the emergence of new exploitable security design flaws in smart contracts.

\section{Related Work}
\label{section:related}
We discuss closely related work, covering smart contract vulnerabilities; empirical evaluations of automated analysis tools that are relevant to our work; and security technical debt.

\textbf{Smart Contract Vulnerabilities.}
Previous studies have illustrated, discussed, or surveyed various security vulnerabilities in blockchain and smart contracts. Li et al. \cite{li2020survey} reviewed the security vulnerabilities in blockchain, the corresponding attacks, and suggested several security solutions, without differentiating between Bitcoin and Ethereum. Similarly, the authors in \cite{hasanova2019survey}, classified the security vulnerabilities based on blockchain generation. They also provided a detailed explanation of known vulnerabilities and subsequent potential attacks. Unlike these studies, we focus on design vulnerabilities in smart contracts run on Ethereum. Other studies such as that of Atzei et al. \cite{atzei2016survey}, discussed 12 known vulnerabilities in Ethereum smart contracts, and classified them into three categories based on the level where they presented: the Solidity level, the EVM level, and the blockchain level. Similarly, authors in \cite{chen2020survey} provided the same classification of vulnerabilities in Ethereum smart contracts; however, they provided a more comprehensive list. In contrast, we aggregated vulnerabilities caused by flaws in a contract’s architectural design, mapped them to their related CWE entries, and classified them based on their impact. 

\textbf{Automated Analysis Tools.}
As several automated tools have recently emerged  to analyse the security of smart contracts, empirical studies have been conducted to compare their real capabilities and the techniques used. Durieux et al. \cite{durieux2020empirical} carried out a systematic evaluation of several state-of-the-art automated tools and discussed their accuracy and efficiency. We also followed a systematic method when it came to selecting the 9 tools that were used in the vulnerabilities identification process. But different from other studies, we identified and analysed a set of tools able to discover a subset of design vulnerabilities. Parizi et al. \cite{10.5555/3291291.3291303} performed an assessment of four static analysis tools with regard to 10 vulnerable smart contracts. In contrast, in our analysis, besides using static analysis tools, we included tools that use dynamic or fuzzing techniques. Additionally, we analysed their ability to identify design vulnerabilities on 16 vulnerable contracts. Leid et al. \cite{leid2020testing} compared the effectiveness of symbolic and fuzzy testing tools by evaluating their effectiveness when analysing smart contracts. Despite a large number of available automated analysis tools, they only considered three tools in their assessment.  

\textbf{Security Technical Debt.}
The nature of work presented in this paper is generally related to applying technical debt to raising the visibility of security risks, and the costly consequences of deploying vulnerable smart contracts. Despite the vast contributions on technical debt in software, only few studies have looked at security-related debts including \cite{izurieta2018position} \cite{izurieta2019leveraging} \cite{nord2016can} and \cite{rindell2019managing}.
In particular, Izurieta et al. \cite{izurieta2018position} established an approach to prioritise security technical debt in a software system. This was related to CWEs entries by leveraging the CWSS scoring systems. Unlike their approach, our study proposes a mechanism to estimate the accumulated debt interest, not only by using the CWSS score, but also by including the activity level and lifespan of the smart contract under consideration. Additionally, we quantify the debt principle of refactoring the vulnerable contract. In \cite{izurieta2019leveraging}, authors mapped attack tactics to the related posed consequences of the exploitable vulnerabilities. This helps to prioritise vulnerabilities that require the most attention to reduce technical debt. In \cite{nord2016can}, the authors applied a preliminary experiment to show the correlation between technical debt and software vulnerabilities. A recent study \cite{rindell2019managing}, regarding security debt, discussed the concept of managing security risk by leveraging technical debt. They emphasised that the combination of software security engineering techniques and technical debt increases the security of software systems. Our work goes beyond existing work on security technical debt and is the first to explicate security technical debts in smart contracts.

\section{Conclusion}
\label{section:conclusion}
In this article, we have presented a debt-aware approach for assessing security design vulnerabilities in smart contracts. We used nine state-of-the-art tools to widen the detection abilities of security design issues in smart contracts. We use CWE catalogue when analysing the identified vulnerabilities and their weaknesses. We adapted the community-informed scoring mechanism to consider contract activity level and the contract lifespan. The combination helps security software engineers to estimate the accumulated debt interests related to design vulnerabilities in a contract. Debt principal was quantified by calculating the gas fee required to redeploy the patched version of a vulnerable contract. Experiment results demonstrated that our approach can allow developers to visualise and prioritise technical debts, rooted in unaddressed smart contract design vulnerabilities. The approach can increase the visibility of debts and their ramifications. 

In future work, we seek to automate the estimation steps of our approach to support faster and more efficient analysis. Automated support will ensure consistency in the analysis and allow security software engineers to quickly estimate the implications of applying alternative security options when designing smart contracts. 


\bibliographystyle{abbrv}
\bibliography{smartdebt}

\begin{thebibliography}{10}

\bibitem{atzei2016survey}
N.~Atzei, M.~Bartoletti, and T.~Cimoli.
\newblock A survey of attacks on ethereum smart contracts.
\newblock {\em IACR Cryptol. ePrint Arch.}, 2016:1007, 2016.

\bibitem{chen2020survey}
H.~Chen, M.~Pendleton, L.~Njilla, and S.~Xu.
\newblock A survey on ethereum systems security: Vulnerabilities, attacks, and
  defenses.
\newblock {\em ACM Computing Surveys (CSUR)}, 53(3):1--43, 2020.

\bibitem{mythx}
CONSENSYS.
\newblock mythx.
\newblock https://mythx.io/.

\bibitem{Ethlint}
R.~Dua.
\newblock Protofire.
\newblock https://github.com/duaraghav8/Ethlint.

\bibitem{durieux2020empirical}
T.~Durieux, J.~F. Ferreira, R.~Abreu, and P.~Cruz.
\newblock Empirical review of automated analysis tools on 47,587 ethereum smart
  contracts.
\newblock In {\em Proceedings of the ACM/IEEE 42nd International Conference on
  Software Engineering}, pages 530--541, Seoul South Korea, 2020. ACM/IEEE.

\bibitem{8823898}
J.~{Feist}, G.~{Grieco}, and A.~{Groce}.
\newblock Slither: A static analysis framework for smart contracts.
\newblock In {\em 2019 IEEE/ACM 2nd International Workshop on Emerging Trends
  in Software Engineering for Blockchain (WETSEB)}, pages 8--15, Montreal, QC,
  Canada, Canada, 2019. ACM/IEEE.

\bibitem{gatteschi2018blockchain}
V.~Gatteschi, F.~Lamberti, C.~Demartini, C.~Pranteda, and V.~Santamar{\'\i}a.
\newblock Blockchain and smart contracts for insurance: Is the technology
  mature enough?
\newblock {\em Future Internet}, 10(2):20, 2018.

\bibitem{githubwiki}
github wiki.
\newblock List of security vulnerabilities, Apr. 2019.

\bibitem{DASP}
N.~Group.
\newblock Decentralized application security project.
\newblock https://dasp.co/.

\bibitem{hasanova2019survey}
H.~Hasanova, U.-j. Baek, M.-g. Shin, K.~Cho, and M.-S. Kim.
\newblock A survey on blockchain cybersecurity vulnerabilities and possible
  countermeasures.
\newblock {\em International Journal of Network Management}, 29(2), 2019.

\bibitem{izurieta2018position}
C.~Izurieta, K.~Kimball, D.~Rice, and T.~Valentien.
\newblock A position study to investigate technical debt associated with
  security weaknesses.
\newblock In {\em 2018 IEEE/ACM International Conference on Technical Debt
  (TechDebt)}, pages 138--142. IEEE, 2018.

\bibitem{izurieta2019leveraging}
C.~Izurieta and M.~Prouty.
\newblock Leveraging secdevops to tackle the technical debt associated with
  cybersecurity attack tactics.
\newblock In {\em 2019 IEEE/ACM International Conference on Technical Debt
  (TechDebt)}. IEEE, 2019.

\bibitem{kruchten2012technical}
P.~Kruchten, R.~L. Nord, and I.~Ozkaya.
\newblock Technical debt: From metaphor to theory and practice.
\newblock {\em Ieee software}, 29(6):18--21, 2012.

\bibitem{leid2020testing}
A.~Leid, B.~van~der Merwe, and W.~Visser.
\newblock Testing ethereum smart contracts: A comparison of symbolic analysis
  and fuzz testing tools.
\newblock In {\em Conference of the South African Institute of Computer
  Scientists and Information Technologists 2020}, pages 35--43, 2020.

\bibitem{li2020survey}
X.~Li, P.~Jiang, T.~Chen, X.~Luo, and Q.~Wen.
\newblock A survey on the security of blockchain systems.
\newblock {\em Future Generation Computer Systems}, 107, 2020.

\bibitem{lohr2020maintenance}
M.~Lohr and S.~Peldszus.
\newblock Maintenance of long-living smart contracts.
\newblock In {\em Software Engineering (Workshops)}, Innsbruck, Österreich,
  2020. ceurws.

\bibitem{luu2016making}
L.~Luu, D.-H. Chu, H.~Olickel, P.~Saxena, and A.~Hobor.
\newblock Making smart contracts smarter.
\newblock In {\em Proceedings of the 2016 ACM SIGSAC conference on computer and
  communications security}, pages 254--269, Vienna Austria, 2016. ACM.

\bibitem{solidity-security-blog}
A.~Manning.
\newblock solidity-security-blog.
\newblock https://bit.ly/2Oi2xsa.

\bibitem{debtexchanges}
C.~Mera-G{\'o}mez, F.~Ram{\'i}rez, R.~Bahsoon, and R.~Buyya.
\newblock A multi-agent elasticity management based on multi-tenant debt
  exchanges.
\newblock In {\em Proceedings of the 12th IEEE International Conference on
  Self-Adaptive and Self-Organizing Systems (SASO 2018)}. IEEE, 2018.

\bibitem{CWE}
MITRE.
\newblock Common weakness enumeration.
\newblock https://cwe.mitre.org/.

\bibitem{CWSS_new}
MITRE.
\newblock Common weakness scoring system.
\newblock https://cwe.mitre.org/cwss/.

\bibitem{manticore}
M.~Mossberg.
\newblock manticore.
\newblock https://github.com/trailofbits/manticore.

\bibitem{mueller2018smashing}
B.~Mueller.
\newblock Smashing ethereum smart contracts for fun and real profit.
\newblock {\em HITB SECCONF Amsterdam}, 9:54, 2018.

\bibitem{10.1145/3377811.3380334}
T.~D. Nguyen, L.~H. Pham, J.~Sun, Y.~Lin, and Q.~T. Minh.
\newblock Sfuzz: An efficient adaptive fuzzer for solidity smart contracts.
\newblock In {\em Proceedings of the ACM/IEEE 42nd International Conference on
  Software Engineering}, ICSE '20, page 778–788, New York, NY, USA, 2020.
  Association for Computing Machinery.

\bibitem{nord2016can}
R.~L. Nord, I.~Ozkaya, E.~J. Schwartz, F.~Shull, and R.~Kazman.
\newblock Can knowledge of technical debt help identify software
  vulnerabilities?
\newblock In {\em 9th Workshop on Cyber Security Experimentation and Test
  ($\{$CSET$\}$ 16)}, 2016.

\bibitem{not-so-smart-contracts}
T.~of~Bits.
\newblock (not so) smart contracts.
\newblock https://bit.ly/2ZX2VyK.

\bibitem{StateoftheDApps}
S.~of~the DApps.
\newblock stateofthedapps.
\newblock https://www.stateofthedapps.com/.

\bibitem{oliva2020exploratory}
G.~A. Oliva, A.~E. Hassan, and Z.~M.~J. Jiang.
\newblock An exploratory study of smart contracts in the ethereum blockchain
  platform.
\newblock {\em Empirical Software Engineering}, pages 1--41, 2020.

\bibitem{openzeppelinforum}
f.~openzeppelin.
\newblock Discuss about smart contract security patterns, vulnerabilities,
  hacks and best practices, Nov. 2020.

\bibitem{ethernaut}
openzeppelin team.
\newblock ethernaut.
\newblock https://ethernaut.openzeppelin.com/.

\bibitem{owasp}
OWASP.
\newblock Open web application security project.
\newblock https://bit.ly/30aNtzc.

\bibitem{10.5555/3291291.3291303}
R.~M. Parizi, A.~Dehghantanha, K.-K.~R. Choo, and A.~Singh.
\newblock Empirical vulnerability analysis of automated smart contracts
  security testing on blockchains.
\newblock In {\em Proceedings of the 28th Annual International Conference on
  Computer Science and Software Engineering}, CASCON '18, page 103–113, USA,
  2018. IBM Corp.

\bibitem{solhint}
Protofire.
\newblock solhint.
\newblock https://github.com/protofire/solhint.

\bibitem{YosRiady}
Y.~Riady.
\newblock Common smart contract vulnerabilities and how to mitigate them, 2018.

\bibitem{rindell2019managing}
K.~Rindell, K.~Bernsmed, and M.~G. Jaatun.
\newblock Managing security in software: Or: How i learned to stop worrying and
  manage the security technical debt.
\newblock In {\em Proceedings of the 14th International Conference on
  Availability, Reliability and Security}, pages 1--8, 2019.

\bibitem{santos2017catalog}
J.~C. Santos, K.~Tarrit, and M.~Mirakhorli.
\newblock A catalog of security architecture weaknesses.
\newblock In {\em 2017 IEEE International Conference on Software Architecture
  Workshops (ICSAW)}, pages 220--223. IEEE, 2017.

\bibitem{blog.positive}
I.~Security.
\newblock blog.positive.
\newblock https://blog.positive.com/.

\bibitem{SWC}
SmartContractSecurity.
\newblock Swc registry.
\newblock https://cwe.mitre.org/cwss/.

\bibitem{sousa2018identifying}
L.~Sousa, A.~Oliveira, W.~Oizumi, S.~Barbosa, A.~Garcia, J.~Lee, M.~Kalinowski,
  R.~de~Mello, B.~Fonseca, R.~Oliveira, et~al.
\newblock Identifying design problems in the source code: A grounded theory.
\newblock In {\em Proceedings of the 40th International Conference on Software
  Engineering}, 2018.

\bibitem{tikhomirov2018smartcheck}
S.~Tikhomirov, E.~Voskresenskaya, I.~Ivanitskiy, R.~Takhaviev, E.~Marchenko,
  and Y.~Alexandrov.
\newblock Smartcheck: Static analysis of ethereum smart contracts.
\newblock In {\em Proceedings of the 1st International Workshop on Emerging
  Trends in Software Engineering for Blockchain}, pages 9--16, Gothenburg
  Sweden, 2018. ACM/IEEE.

\bibitem{tom2013exploration}
E.~Tom, A.~Aurum, and R.~Vidgen.
\newblock An exploration of technical debt.
\newblock {\em Journal of Systems and Software}, 86(6):1498--1516, 2013.

\bibitem{10.1145/3243734.3243780}
P.~Tsankov, A.~Dan, D.~Drachsler-Cohen, A.~Gervais, F.~B\"{u}nzli, and
  M.~Vechev.
\newblock Securify: Practical security analysis of smart contracts.
\newblock In {\em 18 Proceedings of the 2018 ACM SIGSAC Conference on Computer
  and Communications Security}, CCS '18, page 67–82, New York, NY, USA, 2018.
  Association for Computing Machinery.

\bibitem{wang2019blockchain}
S.~Wang, L.~Ouyang, Y.~Yuan, X.~Ni, X.~Han, and F.-Y. Wang.
\newblock Blockchain-enabled smart contracts: architecture, applications, and
  future trends.
\newblock {\em IEEE Transactions on Systems, Man, and Cybernetics: Systems},
  49(11):2266--2277, 2019.

\bibitem{wood2014ethereum}
G.~Wood et~al.
\newblock Ethereum: A secure decentralised generalised transaction ledger.
\newblock {\em Ethereum project yellow paper}, 151(2014):1--32, 2014.

\bibitem{yavo2016}
U.~Yavo.
\newblock Design vulnerabilities: They hide and you can’t catch them.
\newblock https://bit.ly/37S4Bhl.

\bibitem{yu2020smart}
X.~L. Yu, O.~Al-Bataineh, D.~Lo, and A.~Roychoudhury.
\newblock Smart contract repair.
\newblock {\em ACM Transactions on Software Engineering and Methodology
  (TOSEM)}, 29(4):1--32, 2020.

\bibitem{mediumblog}
K.~Zipfel.
\newblock New smart contract weakness: Hash collisions with multiple variable
  length arguments, Jan. 2019.

\end{thebibliography}

\end{document}